\documentclass[singlecolumn,showpacs,secnumarabic,amssymb,superscriptaddress,nobibnotes,aps,pre]{revtex4-2}
\usepackage{amsmath}
\usepackage{amssymb,gensymb}
\usepackage{graphicx}
\usepackage{url,hyperref}
\usepackage[usenames,dvipsnames]{xcolor}
\hypersetup{colorlinks=true, linkcolor=BrickRed, urlcolor=blue!50!black, citecolor=blue!50!black}
\usepackage[capitalize]{cleveref}
\usepackage{cleveref}

\AtBeginDocument{}
\crefrangelabelformat{equation}{(#3#1#4)$-$(#5#2#6)}
\begin{document}
\title{Mpemba Effect and Superuniversality across Orders of Magnetic Phase Transition}
\author{Sohini Chatterjee}
\affiliation{Theoretical Sciences Unit and School of Advanced Materials, 
Jawaharlal Nehru Centre for Advanced Scientific Research, Jakkur P.O., 
Bangalore 560064, India}
\author{Soumik Ghosh}
\affiliation{Theoretical Sciences Unit and School of Advanced Materials, 
Jawaharlal Nehru Centre for Advanced Scientific Research, Jakkur P.O., 
Bangalore 560064, India}
\author{Nalina Vadakkayil}
\affiliation{Theoretical Sciences Unit and School of Advanced Materials, 
Jawaharlal Nehru Centre for Advanced Scientific Research, Jakkur P.O., 
Bangalore 560064, India}
\affiliation{Complex Systems and Statistical Mechanics, Department of Physics and Materials Science, University of Luxembourg, Luxembourg, L-1511, Luxembourg}
\author{Tanay Paul}
\affiliation{Theoretical Sciences Unit and School of Advanced Materials, 
Jawaharlal Nehru Centre for Advanced Scientific Research, Jakkur P.O., 
Bangalore 560064, India}
\author{Sanat K. Singha}
\affiliation{Theoretical Sciences Unit and School of Advanced Materials, 
Jawaharlal Nehru Centre for Advanced Scientific Research, Jakkur P.O., 
Bangalore 560064, India}
\affiliation{Assam Energy Institute, A Centre of Rajiv Gandhi Institute of 
Petroleum Technology, Sivasagar 785697, India}
\author{Subir K. Das}
\email{das@jncasr.ac.in}
\affiliation{Theoretical Sciences Unit and School of Advanced Materials, 
Jawaharlal Nehru Centre for Advanced Scientific Research, Jakkur P.O., 
Bangalore 560064, India}
\date{\today}

\begin{abstract}
The quicker freezing of hotter water, than a colder sample, when 
quenched to a common lower temperature, is referred to as the 
Mpemba effect (ME). While this counter-intuitive fact remains a 
surprize since long, efforts have begun to identify similar effect 
in other systems. Here we investigate the ME in a rather general context
concerning magnetic phase transitions. From Monte Carlo simulations 
of model systems, viz., the $q$-state Potts model and the Ising model,
with varying range of interaction and space dimension, we assert 
that hotter paramagnets undergo ferromagnetic ordering faster than the 
colder ones. The above conclusion we have arrived at following the 
analyses of the simulation results on decay of energy and growth in ordering 
following quenches from different starting temperatures, to fixed final 
temperatures below the Curie points. We have obtained a unique scaling picture, 
on the strength of the effect, with respect to the variation in 
spatial correlation in the initial states. These results are valid irrespective 
of the order of transition and relevant to the understanding of 
ME in other systems, including water. 
 \end{abstract}

\maketitle

\section{Introduction}

If two bodies of liquid water, differing in temperature, 
are placed in contact with a thermal reservoir, operating at a subzero 
temperature ($<0\degree$C), the most common prediction will be that the colder between 
the two will freeze faster. The report in Ref. \cite{mpemba}, by Mpemba 
and Osborne, however, contradicts such an expectation. There has been a 
surge \cite{mpemba,skd,bechhoffer,auerbach,jeng,xi,jin,tao,tang,avinash,ahn,
greaney,chaddah,lasanta,torrente,rajesh,gomez,baity,lu_raz,gal_raz,nv,oraz2,
biswas,walker,vynnycky,yang,lowen,cao,first_order} in interest 
in further investigating this forgotten counter-intuitive 
fact, which found mention even in the works of Aristotle \cite{aristotle},
now referred to as the Mpemba effect (ME).
In recent times, questions relevant to the ME are posed in more general ways \cite{skd}, 
going much beyond the domain of water 
\cite{mpemba,skd,bechhoffer,auerbach,jeng,xi,jin,tao,tang,avinash,ahn}: When two samples 
of the same material, from two different temperatures, are quenched to a common
lower temperature, which one will reach the new equilibrium quicker? If there 
exists a point of phase transition, between the initial and the final 
temperatures, for which starting temperature the transformation will 
occur earlier? There is a rapid growth in interest in studying 
systems of different types. Experimental studies on systems such as 
colloidal systems \cite{avinash}, 
clathrate hydrates \cite{ahn}, carbon nanotube resonators \cite{greaney} 
and magnetic alloys \cite{chaddah}
show the presence of ME. In the theoretical and computational literature, 
the varieties that are studied include 
granular matter \cite{lasanta,torrente,rajesh,gomez}, spin glass 
\cite{baity} and few other systems of magnetic origin \cite{lu_raz,gal_raz,nv,oraz2}. 
Nevertheless, the underlying 
reason remains a puzzle, inviting the need for stronger theoretical intervention. 
The pertaining new questions are fundamental from statistical mechanical 
and other theoretical points of view. In addition, the effect can be 
exploited to much practical advantages \cite{cao}.

In the case of water, for a transformation from liquid to solid, overcoming metastability 
is a serious problem \cite{metastability}. Nucleation there, as well as in general, is strongly 
influenced by the choice of the final temperature ($T_f$), the type and the volume fraction of 
impurity, etc. However, how the role of the starting temperature ($T_s$) enters the picture 
is an important new fundamental issue. Is it that the above mentioned metastable 
aspect gets affected by the nature of the initial thermodynamic state in an unexpected 
order? With in-built frustration in the model Hamiltonian, some of the theoretical works 
perhaps set the objective of exploring this angle. Such studies 
are by primarily involving the magnetic systems with anti-ferromagnetic interactions, 
including spin glasses. However, simpler systems must also be considered. 
If the effect is found to be present in their evolution dynamics, route to 
a proper understanding can be easier.

In fact, the standard ferromagnetic Ising model, without any impurity, is seen 
to exhibit the effect \cite{nv}. The positive observation of the ME there was attributed 
to the structural changes associated with the critical divergence of the spatial correlation 
with the variation of $T_s$ in the paramagnetic phase \cite{skd,nv}. To justify the 
validity of the attribution and estimate the corresponding quantitative influence, it is important to 
study the model in other situations like in different space dimensions ($d$) and with 
varying range of interactions. Very importantly, it should be checked what 
are the effects of the order of transitions \cite{skd,first_order,oraz2}. 
This is by considering the fact that in the case of water, the transition 
is of first order character, while for the Ising 
model the problem was designed \cite{nv} in such a way that the influence of second 
order transition is captured. Keeping these in mind \cite{skd,first_order,oraz2}, here we chose the $q$-state 
Potts model \cite{landau,potts_binder} for which the order of transition changes with the variation of its 
number of states $q$. Interestingly, we observe the presence of the ME in all the 
above mentioned cases. 
For the Potts model, it appears that with the increase of $q$ the effect 
gets weaker, as far as the variation in $T_s$ is concerned. However, 
interestingly, regarding the effects of spatial correlation in the initial states
there exists a unique scaling, not only for the simple variation of $q$, 
but also with the change of the order of transition and space dimension. We believe, in addition to being 
important for the class of systems considered, this study also provides 
crucial angle with respect to the interpretation of the effect in water.

\section{Models and Methods}

As already mentioned, here we study a class of discrete spin systems 
with pure ferromagnetic interactions. We investigate the Ising model \cite{landau}
in different dimensions, having short and long-range inter-site potentials \cite{lr_bray_pre}.
Furthermore, generalization of this two-component system into multi-component ones 
have also been considered. In general, the Hamiltonian can be written as
$H = -\sum _{i (\neq) j} J(r_{ij}) \delta_{S_i,S_j};\,\, 
 S_i,S_j = 1,2, ... , q;$ $r_{ij}$ being the separation between lattice sites.
For $q=2$, this Potts model Hamiltonian
corresponds to the Ising model, differing only by a factor of $2$. The latter, by 
correcting for the factor, we have studied in $d=2$ and $3$, with nearest neighbor (NN)
interactions, by setting the interaction strength $J$ to unity, on square and simple 
cubic lattices, respectively. In $d=2$, we have also presented results for the 
long-range (LR) version of the Ising model with \cite{lr_bray_pre}
$J(r_{ij}) = 1/r_{ij}^{2+\sigma}$, for $\sigma=0.8$, again using the square lattice. Most extensive results
are obtained for the (short-range) Potts model, $q$ varying between $2$ and $10$.
The critical temperature for this model has the $q$-dependence \cite{landau}
$T_c = J/[k_B \ln(1+\sqrt{q})]$, $k_B$ being the Boltzmann constant, to be set to unity. 
Depending on the value of $q$ the order of transition can alter. 
For $q > 4$ the model loses its ``critical" character \cite{potts_binder}, the
transition being of first order. 
For the Ising case, the values of $T_c$ in $d = 2$ and $3$ are $\simeq 2.269\,J/k_B$
and $\simeq 4.51\,J/k_B$, respectively \cite{landau}. For the long-range case we have 
used \cite{horita} $T_c = 9.765\,J/k_B$.

The kinetics of transition from para to ferro states are studied
via Monte Carlo simulations \cite{landau}, with the Glauber spin-turn 
mechanism \cite{landau}. The preparation 
of the initial configurations near the critical point encounters critical slowing down \cite{landau}. 
To avoid this, we have used cluster algorithms. 
In the case of Ising or Potts model, this is done by implementing the 
Wolff algorithm \cite{wolff}, and for the LR Ising model, we have used the Fukui-Todo 
algorithm \cite{fukui,horita}. The presence of the correlated spatial fluctuations 
in a system and its variation 
with temperature can be quantified via the calculation of the structure factor: 
$S(k,t) = \langle\psi_k\,(t)\,\psi_{-k}(t)\rangle,$
$\psi_k\,(t)$ being the Fourier transform of the order parameter \cite{op_sp} $\psi\,(\vec{r},t)$ $(= \exp(i
\,\theta(\vec{r})); \:\:
\theta = 2\pi n/q, \:\: n = 1,\dots,q$). 
In the small $k$ regime [$\in (0,0.4)$ or shorter], for the short-range cases, $S(k,t)$ is described well 
by the Ornstein-Zernike relation \cite{fisher,stanley},
    $S(k) = k_B \:T_s\:\chi/(1 + k^2 \: \xi^2),$
$\chi$ being the susceptibility and $\xi$ the correlation length. 
The average domain length, $\ell(t)$, of the clusters, during an 
evolution towards the ferromagnetic state, 
has been estimated via the first moment of the domain size 
distribution function \cite{sm_domain},
$P(\ell_d)$, i.e., $\ell(t)=\int P(\ell_d,t)\ell_d d\ell_d$,  where $\ell_d$ is 
the distance between two consecutive interfaces along a given direction. 

\section{Results}

\begin{figure}
\centering
\includegraphics*[width=0.9\textwidth]{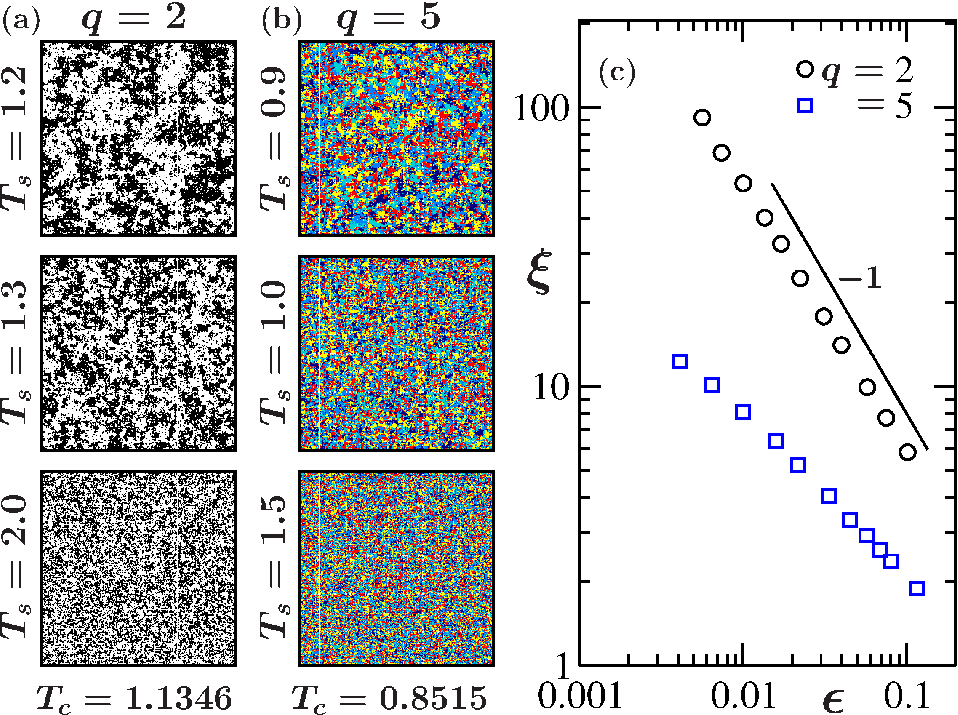}
\caption{Typical equilibrium configurations, for $q=2$ and $5$ state Potts models,
are shown in (a) and (b) from different starting temperatures $T_s$ that are located 
above the respective critical temperatures $T_c$. (c) The plots of the correlation 
lengths, $\xi$, versus $\epsilon$ ($=(T_s-T_c)/T_c$), for the same Potts models. 
For $q=2$, $\xi$ diverges as $\epsilon^{-\nu}$, with $\nu=1$ (see the solid line). 
These results are obtained for $L=256$.}
\label{fig_snap}
\end{figure}

In Fig. \ref{fig_snap} we depict how the choice of $T_s$, in the case of Potts model, 
can influence the structural features in the initial configurations, for different values of $q$.
For both the considered cases, viz., $q=2$ and $5$, the configurations at higher $T_s$ 
[see the snapshots at the bottom of the columns in (a) and (b)] appear
random or structureless. With the decrease of $T_s$, spatial correlations emerge. This is 
more clearly identifiable in the case of $q=2$ for which one expects the critical
divergence \cite{fisher,stanley} $\xi \sim \epsilon^{-\nu}$, with $\nu = 1$. For a wide range of 
$\epsilon$ ($= |T_s-T_c|/T_c$) such a behavior can be appreciated from 
Fig. \ref{fig_snap}(c). For $q=5$, the phase transition is of first order \cite{potts_binder}, 
and we do not associate any exponent with the data set.
The enhancement in the value of $\xi$ can be appreciated for this $q$ as well.
In this case, the bending on the log-log scale over the whole presented range 
can well be, in addition to the finite-size effects, due to the first-order nature of the transition. 
Our expectation is that for this $q$ the effect will be weaker. We proceed with the 
objective of quantifying it and to investigate if there exists any scaling rule \cite{skd} for 
arbitrary $q$ that can as well comply with the other considered models. 

\begin{figure}
\centering
\includegraphics*[width=0.9\textwidth]{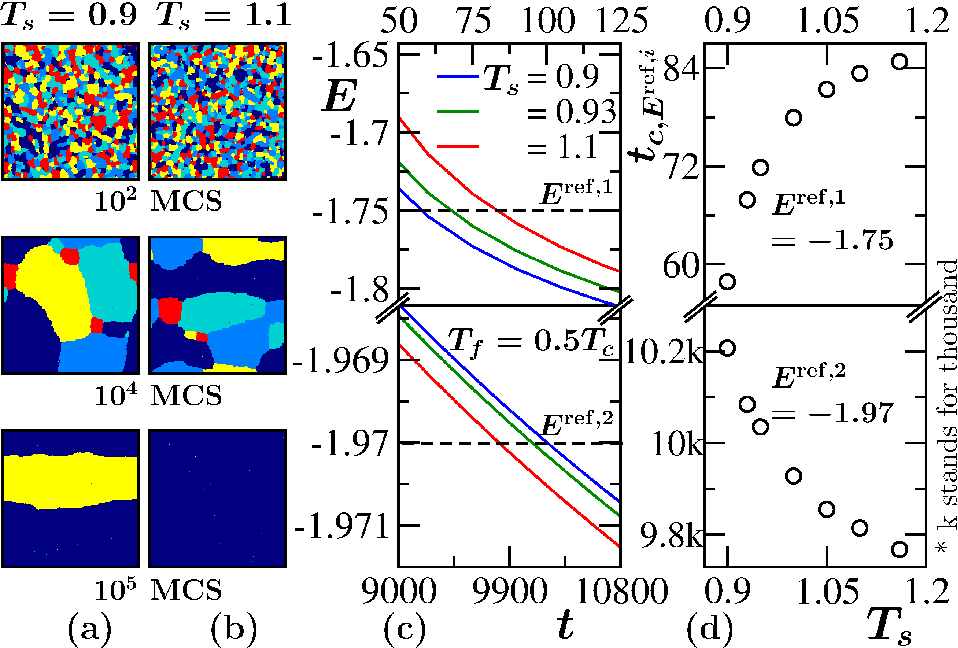}
\caption{Demonstration of relaxation in the 5-state 
Potts model, following quenches to $T_f = 0.5\,T_c$, 
from different $T_s$ values, with equal fraction of different spin states in boxes having $L=256$. 
In (a) and (b) we show evolution snapshots, taken at different times, 
in units of standard Monte Carlo steps, for the systems 
initially at (a) $T_s = 0.9$ and (b) $T_s = 1.1$. 
Different colours represent $q$ different Potts states. (c) Plots of energy versus time, 
following quenches from several $T_s$ values. The upper 
half corresponds to early time behavior and the 
lower one shows the late time behavior. 
The dashed horizontal lines are drawn to extract 
the times, $t_{c,E^{{\rm ref},i}}, i =1,2$, corresponding to the crossings of 
certain values of energy $E^{{\rm ref},i}$ by the energy curves of the systems 
with different $T_s$. 
(d) Plots of $t_{c,E^{{\rm ref},1}}$ (upper panel) and $t_{c,E^{{\rm ref},2}}$ (lower panel), versus $T_s$.}
\label{fig_pottsE}
\end{figure}

We quench the systems from different \cite{chak_das} $T_s$ to $T_f = 0.5\,T_c$, for a 
large set of $q$ values. In Fig. \ref{fig_pottsE} we 
show results obtained during evolutions following such quenches, for the $5$-state Potts model. 
In parts (a) and (b) we show the snapshots at different stages of 
evolutions for $T_s=0.9$ and $1.1$, respectively.
It is  evident that the system from the higher $T_s$ 
reaches the final equilibrium faster. This comparative picture is true not only for the chosen set
of initial configurations, the observation indeed stands correct for a vast majority of the 
combinations of starting configurations.
In part (c) we look at the decay of the average energy ($E$) of the systems during the relaxation processes.
We have included results for several $T_s$. 
Each of the data sets is presented after averaging over $300000$ 
independent initial configurations. For clarity, we have enlarged
the early and late time behavior separately, in the upper 
and lower panels, respectively. The orders of appearances of the plots, in terms of $T_s$,
are systematic and opposite in the two panels.
This implies that there are crossings amongst the energy curves, due to faster equilibrations of 
configurations prepared at higher $T_s$. This is the essence of Mpemba effect \cite{mpemba,skd,baity,nv}. For 
a demonstration of systematicity, we perform the following exercise.
The dashed horizontal lines in these panels correspond 
to two reference energy values, $E^{{\rm ref},1}$ and $E^{{\rm ref},2}$. We 
calculate the crossing time 
between a dashed line and the energy curve of the systems starting from each of the $T_s$ values. Such a 
crossing time is denoted by $t_{c,E^{{\rm ref},i}}, i=1,2$. 
Part (d) of Fig. \ref{fig_pottsE} 
shows $t_{c,E^{{\rm ref},i}}$ as a function of $T_s$. 
The upper and lower panels capture the early and late time 
behavior, respectively. The early time quantity, i.e., $t_{c,E^{{\rm ref},1}}$, increases 
with the increase 
in $T_s$ but, at late times we see a different behavior, i.e., $t_{c,E^{{\rm ref},2}}$ decreases with 
the increase in $T_s$. This implies faster relaxation of the systems 
with higher $T_s$, indicating the presence of ME.

\begin{figure}
\centering
\vspace{0.2cm}
\includegraphics*[width=0.9\textwidth]{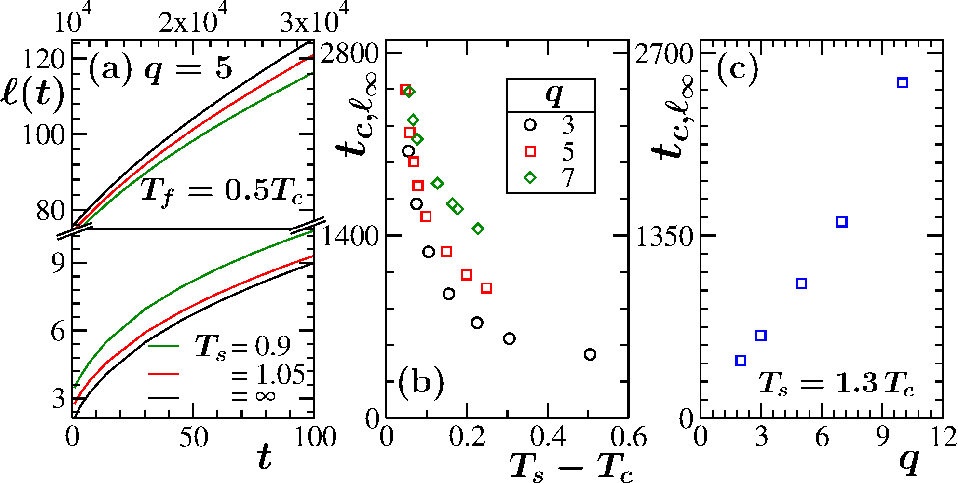}
\caption{(a) The plots of $\ell(t)$ versus $t$, for the 5-state Potts model, 
for quenches from various $T_s$ to $T_f = 0.5\,T_c$. The lower frame captures 
the early time trend, while the upper frame 
depicts the late time behavior. 
(b) Plots of time, $t_{c,\ell_{\infty}}$, corresponding to crossing between growth curves for systems starting at $T_s=\infty$ and a finite $T_s$,
as a function of $T_s -T_c$, for different $q$ values. 
(c) Plot of $t_{c,\ell_{\infty}}$ versus $q$ for 
systems prepared at $T_s = 1.3\,T_c$.
All results correspond to $L = 256$. }
\label{fig_length_potts}
\end{figure}
The faster relaxation of the higher $T_s$ systems
can as well be quantified by calculating the average domain length, 
$\ell(t)$, a key probe to investigate coarsening dynamics 
\cite{sm_domain,chak_das,bray_article}. In Fig. \ref{fig_length_potts}(a), 
we plot $\ell(t)$, vs $t$, for different $T_s$ values, for $q=5$.
The early time behavior for different $T_s$ 
are presented in the lower part of the divided graph. The late time comparisons are in the upper part. 
The systems starting at higher $T_s$ tend to approach the new equilibrium earlier. 
This conveys a picture same as that derived from the energy decay, further strongly suggesting the presence of the Mpemba effect. 
We record the times at 
which the domain lengths of the systems at different finite $T_s$ ($<\infty$) values
are crossed or overtaken by the corresponding plots for the systems starting from 
$T_s = \infty$. We denote this by $t_{c,\ell_{\infty}}$.
In Fig. \ref{fig_length_potts}(b) we have plotted $t_{c,\ell_{\infty}}$ as a function of $T_s-T_c$, for
a few values of $q$, covering transitions of first as well as second order varieties.
For each $q$, the crossing time increases with the approach of $T_s$ to the corresponding $T_c$.
Given that depending upon the value of $q$ the nature of critical fluctuation is different, presence of any 
unique scaling behavior may not emerge from this figure. 
It appears, nevertheless, that for a given distance of $T_s$ 
from $T_c$, the crossing time is longer for higher $q$.
A quantitative picture for this is shown in Fig. \ref{fig_length_potts}(c). This is a signature that the ME 
gets weaker with the increase of $q$. Considering the influence of both $q$ and $T_s$, 
the issue, however, is complex, that we address later.

\begin{figure}
\centering
\includegraphics*[width=0.9\textwidth]{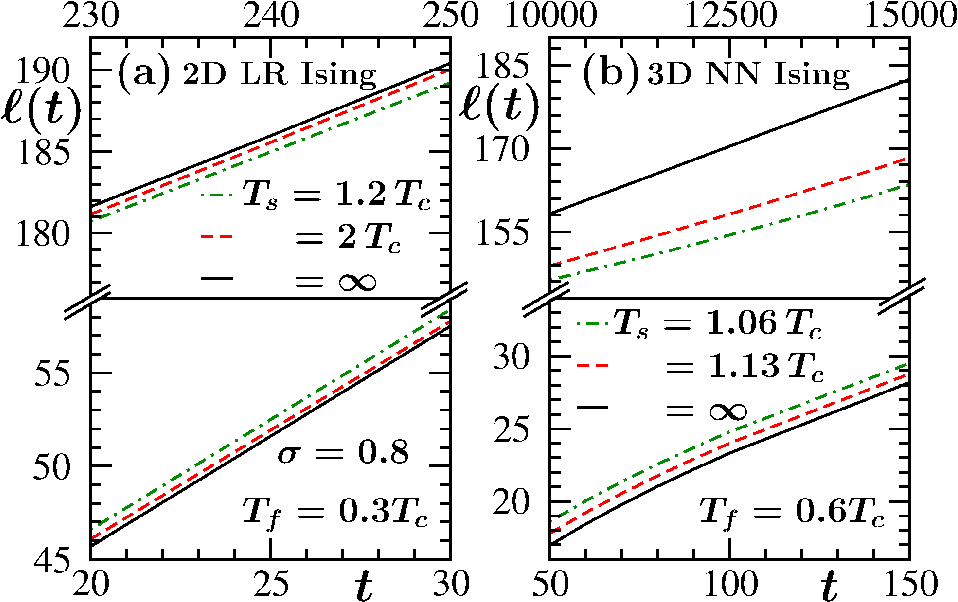}
\caption{(a) Plots of $\ell(t)$ versus $t$, corresponding 
to a few different $T_s$ values, for quenches to $0.3\,T_c$, 
for the LR Ising model. The value of $\sigma$ is $0.8$ 
and we have $L = 1024$. 
(b) Same as (a) but here the results are for the 3D nearest-neighbor (NN) 
Ising model with $T_f = 0.6\,T_c$, and $L=256$.}
\label{fig_LR_l_diffTs}
\end{figure}

Next we check whether the same scenario is true for the case of the LR Ising 
model. Due to the demanding computation, we analyze results for this case after
averaging over $100$ independent initial configurations. 
Note that the LR systems encounter finite-size effects much faster 
than its short-range counterpart, due to faster growth 
\cite{lr_bray_pre,janke,skd_ghosh} with the decrease of $\sigma$. 
To avoid this problem, we choose big systems and a 
large value of $\sigma$, viz., $\sigma=0.8$, which, nevertheless, falls well within 
the long-range interaction domain \cite{lr_bray_pre}. 
In Fig. \ref{fig_LR_l_diffTs}(a) we plot $\ell(t)$, vs $t$, 
for quenches to $T_f = 0.3\,T_c$, from three $T_s$ values, with $L=1024$.
From these plots it is clear that the systems with the highest $T_s$ 
have the largest $\ell(t)$, at late times. 
Thus, ME appears to be present in the LR Ising model as well. 
Note that because of the above mentioned reasons we have used Ewald summation \cite{horita,ewald1}, 
and parallelized our codes, in this case, to speed up the output. 

So far we have dealt with $2\mbox{D}$ systems.
Now we present results from the $3\mbox{D}$ NN
Ising model in Fig. \ref{fig_LR_l_diffTs}(b), where also the
faster relaxation of the systems for the higher $T_s$ value
is quite clear. Here we have quenched the systems from different 
initial $T_s$ values to $T_f = 0.6\,T_c$. These results are presented 
after averaging over runs with 1440 independent initial 
configurations, with $L=256$.

\begin{figure}
\centering
\vspace{0.2cm}
\includegraphics*[width=0.9\textwidth]{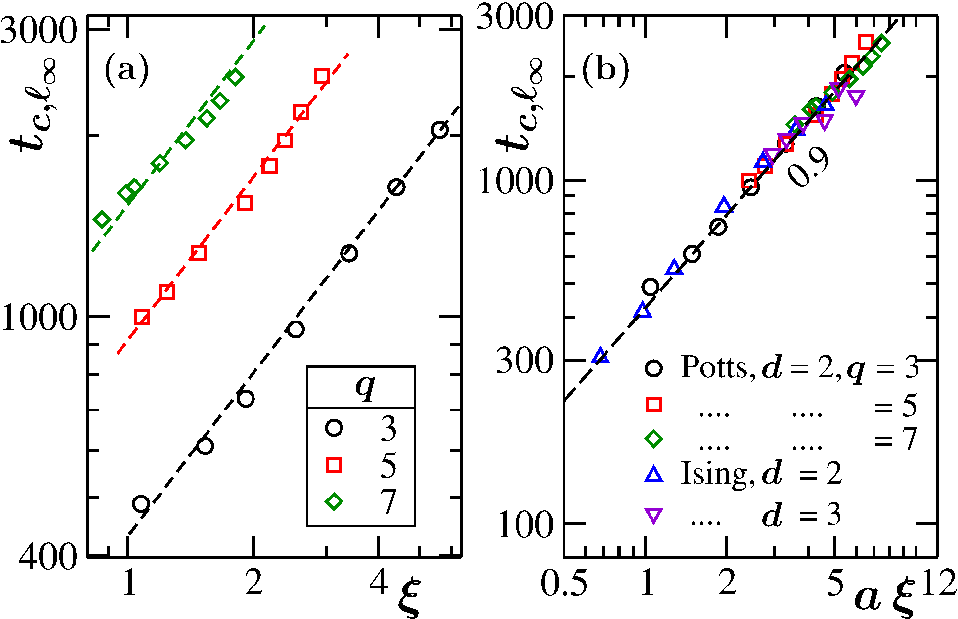}
\caption{(a) Plots of $t_{c,\ell_{\infty}}$ versus $\xi$, for the Potts model, with a few different $q$
values, on a double-log scale. (b) Same as (a) but here the abscissa of the data sets are scaled by constant factors to obtain a ``possible" overlap. In addition to the results from the Potts cases ($q\ge 3$), here we have included data for the NN Ising model, from different space dimensions. We expect discrepancy in scaling very close to $T_c$ due to strong finite-size effects from multiple sources. Dashed lines represent power-laws.}
\label{fig_scalexi}
\end{figure}

Returning to the Potts results in Fig. \ref{fig_length_potts}, we 
recall that an important objective of our work is to 
obtain a scaling picture \cite{skd}. Note that for different $q$ values, 
one expects differing fluctuations in the critical 
vicinity. Thus, a unique behavior of the data sets in 
Fig. \ref{fig_length_potts}(b) should not be expected. It 
is more instructive to replace the abscissa variable there 
by $\xi$. Results from such an exercise is shown
in Fig. \ref{fig_scalexi}(a). On a log-log scale it appears 
that the data sets from different $q$ are reasonably parallel 
to each other. In Fig. \ref{fig_scalexi}(b), thus, we introduce a prefactor 
$a$, for the abscissa, constant for a particular value of $q$, 
to obtain an overlap of the data sets in Fig. \ref{fig_scalexi}(a). A nice collapse 
of the data sets can be appreciated. In fact, the results for the
$2\mbox{D}$ and $3\mbox{D}$ Ising models also comply with that. 
It is worth mentioning here that accurate estimations of 
the crossing times require huge statistics.

\section{Conclusion}

We have investigated the presence of the Mpemba effect \cite{mpemba,skd,bechhoffer,auerbach,jeng} 
during para- to ferromagnetic
transitions in several model systems with discrete spin values.
These include short-range Ising model in $d=2$ and $3$, as well as long-range Ising model in $d=2$.
Very extensive set of results are presented for the $q$-state Potts model for 
a wide range of $q$ values. It is important to note that in none of the considered models there 
exist in-built frustration.
Irrespective of the space dimension, range of interaction and 
order of transition, we have observed the Mpemba effect. It has interesting connection
with the length of spatial correlations at the considered initial temperatures.
The relative delay in approach to the 
final equilibrium, following quenches from para to ferro regions, with the lowering of starting 
temperatures, has unique dependence upon $\xi$. For second order transitions we 
have obtained a universal scaling for models with critical exponent 
$\nu$ varying nearly by a factor of $1.6$. More interestingly, the scaling is valid for 
even first order transitions. 
This implies, for a given model, if two initial temperatures possess nearly same spatial correlations, 
possible for large $q$, configurations from these states will 
equilibrate almost simultaneously at the final temperature, 
showing no detectable ME, even for large differences in the $T_s$, in terms of times taken for
reaching the final destination. We believe that our results contain important 
message for the understanding of the effect in water.
Particularly the observation of it in the cases of first order transition
can shed light on the mystery with respect to the latter. 
It will be interesting to investigate how the power-law dependence upon $\xi$, 
with exponent $0.9$, may be connected to the scaling picture described in Ref. \cite{skd}.

%\section*{Author contributions}

\textbf{Author contributions}: SKD proposed the topic, designed the problem, 
participated in the analyses, 
supervised the work and wrote the manuscript. 
NV oversaw a few coding details and took part in progress on all the models 
at the initial stages, alongside contributing to the writing. 
SC obtained all the final results on the Potts model, analyzed these, and contributed to the writing. 
SG and TP obtained and analyzed the results on the long-range and 
the 3D Ising models, respectively. 
Simulations of SKS provided the first hints of the Mpemba effect in the 3D Ising model.

%\section*{Acknowledgements}

\textbf{Acknowledgments}: SKD acknowledges a discussion with R. Pandit at an early stage and 
partial financial support from Science and Engineering Research Board, India,
via Grant No. MTR/2019/001585. The authors are thankful to the supercomputing facility, 
PARAM Yukti, at JNCASR, under National Supercomputing Mission.

%\section{References}

\end{document}